\documentclass[letter,twocolumn]{jpsj2} 
\usepackage{bm}
 
\title{Emergence of Antiferromagnetic Correlation in LiTi$_{2-x}$V$_{x}$O$_{4}$ via $^{7}$Li NMR}

\author{\textsc
{Yutaka ITOH}\thanks{E-mail: itoh@kuchem.kyoto-u.ac.jp}
, \textsc{Naofumi MORITSU}, and 
\textsc{Kazuyoshi YOSHIMURA}
}

\inst{Department of Chemistry, Graduate School of Science, Kyoto University, Kyoto 606-8502}

\abst{We report $^{7}$Li NMR studies of V-substitution effects on spinel oxide superconductor LiTi$_2$O$_4$ ($T_\mathrm{c}$ = 13.4 K). 
In LiTi$_{2-x}$V$_x$O$_4$ ($x$ = 0 $-$ 0.4),
the V substitution for the Ti site 
suppressed the relative volume fraction of superconductivity faster than $T_\mathrm{c}$. 
From the observation of fairly homogeneous enhancement
in a $^{7}$Li nuclear spin-lattice relaxation rate, 
we conclude that  the V substitution changes electron correlation effects through electron carrier doping from quarter electron filling 3$d^{0.5}$ to 3$d^{1.5}$ and then the antiferromagnetic correlation 
emerges.  
}

\kword{LiTi$_2$O$_4$, NMR, vanadium substitution, antiferromagnetic spin fluctuation}

\begin{document}
\maketitle

Transition-metal spinels have attracted great interests,
because intriguing electronic phases and their competition are associated with 
frustration effects on spin, charge, orbital networks and the coupling with crystal lattice~\cite{JP2007}.
A normal spinel LiT$_2$O$_4$ is a type-II  oxide superconductor with
a high $T_\mathrm{c}$ $\sim$13 K~\cite{Johnston,Mos}.
The crystal structure is stable at low temperatures. 
LiTi$_2$O$_4$ is an $s$-wave superconductor,
because of a Hebel-Slichter peak in $^{7}$Li nuclear spin-lattice relaxation rate~\cite{MItoh} and activation-type specific heat~\cite{Sun}. 
The mystery is the unconventional break down of the superconductivity.      
Although the coherence length $\xi$ is long enough~\cite{Sun},
the superconductivity is suppressed as the decrease in the superconducting volume fraction
for Li-deficient Li$_{1-\delta}$Ti$_2$O$_4$~\cite{Rygula} and Li-rich Li$_{1+y}$Ti$_{2-y}$O$_4$~\cite{Johnston,Ueda}. 
The Li deficiency and Li substitution for Ti site make the conduction band from the quarter filling to
the band insulating states.   

The full solid solution of LiTi$_{2-x}$V$_x$O$_4$ (0$\leq x\leq$ 2) is known~\cite{Tsuda}.
An itinerant-electron spinel LiV$_2$O$_4$ shows a heavy electronic specific heat and a Curie-Weiss spin susceptibility~\cite{Kondo}. 
Although the ferromagnetic spin fluctuations were estimated from NMR data~\cite{Fujiwara}, 
the antiferromagnetic fluctuations were observed at low temperatures by neutron scattering experiments~\cite{ND1,ND2}. 
The spin fluctuation spectrum may not be simple~\cite{ND2,PES}. 
Ti substitution for LiV$_2$O$_4$ induces spin-glass-like behaviors and the microscopic effects were studied~\cite{Trinkl}.  
The microscopic effects of V substitution for LiTi$_2$O$_4$, however, have been poorly understood.

One may expect two electronic effects of V substitution for Ti sites in LiTi$_2$O$_4$. 
One is a simple pair-breaking effect of additional impurity electron spin $\Delta S$ = 1 on the superconductivity. 
The V electrons are assumed to have localized moments. 
The randomly distributed local moments can induce non-exponential NMR relaxation~\cite{Mc,ItohNi,ItohZn,JohnstonNMR}.  
The other is the carrier doping effect on electron correlation through band filling from quarter electron filling 3$d^{0.5}$ to 3$d^{1.5}$. 
The V electrons are assumed to hybridize the Ti conduction electrons and then to be itinerant. 
Even for the random crystalline potentials introduced through the substitution, 
NMR linewidths may be broader but
the NMR relaxation can have a single spin-lattice relaxation time~\cite{ItohSr}. 
 
In this Letter, we report the $^{7}$Li NMR studies of LiTi$_{2-x}$V$_x$O$_4$ ($x$ = 0 $-$ 0.4).  
The V substitution was found to suppress the relative volume fraction of superconductivity 
faster than $T_\mathrm{c}$ and 
to enhance fairly homogeneously and largely the $^{7}$Li nuclear spin-lattice relaxation but not so much the Knight shift.   
The antiferromagnetic correlation was concluded to emerge in the V substituted samples.  

\begin{figure}[bp]
	\begin{center}
		\includegraphics[width=7cm, clip]{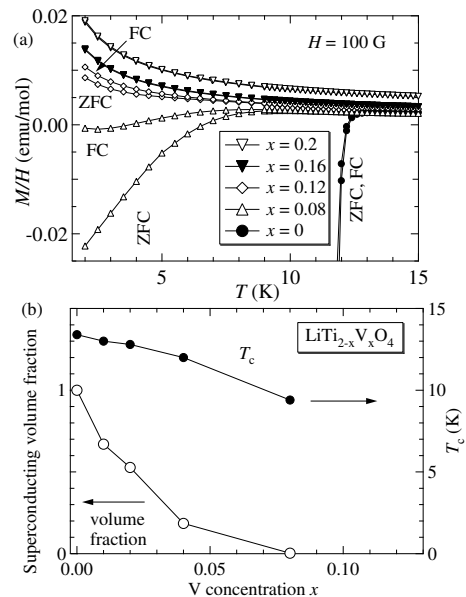}
	\end{center}
	\caption{(a) Uniform magnetic susceptibility $M/H$ at 100 G after cooling in a zero field (ZFC) and in a field of 100 G (FC) for LiTi$_{2-x}$V$_x$O$_4$. 
	(b) V concentration dependence of $T_\mathrm{c}$ and the relative volume fraction of superconductivity to $x$ = 0 for LiTi$_{2-x}$V$_x$O$_4$. The solid curves are visual guides. 
	The bifurcation of temperature hysteresis and the onset of diamagnetic response of $M/H$ were defined as the superconducting transition temperature $T_\mathrm{c}$. 
	}
		\label{fig_Tc}
\end{figure}%
Powder samples of LiTi$_{2-x}$V$_x$O$_4$ were synthesized by a solid state reaction method with a precursor of Li$_{4/3}$Ti$_{5/3}$O$_4$, because Ti is easily oxidized to be Ti$^{4+}$~\cite{Ueda}.  
First, Li$_{4/3}$Ti$_{5/3}$O$_4$ was synthesized from the mixture of preheated dry Li$_2$CO$_3$ (99.9 $\%$) and TiO$_2$ (99.99 $\%$) after ref. 7.     
Next, the mixtures of Li$_{4/3}$Ti$_{5/3}$O$_4$, TiO$_2$, Ti and V$_2$O$_3$ were sealed in evacuated quartz tubes and then fired at 760 $^\circ$C for 0$\leq x \leq$0.04, at 800 $^\circ$C for 0.08$\leq x \leq$0.16, and  at 850 $^\circ$C for 0.2$\leq x \leq$0.4 in a week.
The powder X-ray diffraction patterns indicated all the samples of $x \leq$ 0.04 in a single phase with spinel structure and
the samples of $x \geq$ 0.08 with a small amount of unreacted V$_2$O$_3$.    
The cryopreservation method at 77 K in liquid nitrogen was employed to keep the samples fresh. 

We performed high resolution Fourier-transformed $^{7}$Li (nuclear spin $I$ = 3/2 and nuclear gyromagnetic ratio $\gamma_{\rm n}$/2$\pi$ = 16.546 MHz/T) NMR measurements 
of free induction decay signals or the nuclear spin-echoes at $H$ = 7.48414 T.
The applied magnetic field was estimated from the reference material LiCl$aq$.  
Nuclear spin-lattice relaxation times were measured by an inversion recovery technique. 

\begin{figure}[tp]
	\begin{center}
		\includegraphics[width=6.8cm, clip]{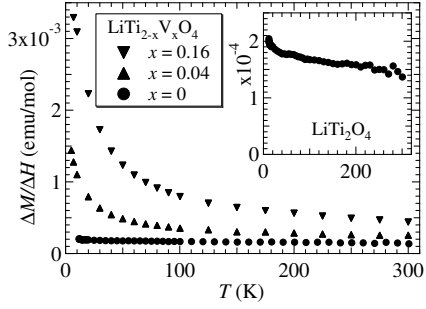}
	\end{center}
	\caption{V substitution effects on temperature dependence of magnetic susceptibility $\chi$ defined by $\Delta M/\Delta H$ at 4 and 5 T for $x$ = 0, 0.04, and 0.16. 
	The inset shows $\Delta M/\Delta H$ for pure LiTi$_2$O$_4$ on an expanded scale.
	}
		\label{fig_XT}
\end{figure}%
Magnetization was measured by a SQUID magnetometer (Quantum Design MPMS)
for $x$ = 0, 0.01, 0.02, 0.04, 0.08, 0.16, 0.2 and 0.4. 
 Figure \ref{fig_Tc} (a) shows uniform magnetic susceptibility $M/H$ at 100 G after cooling in a zero field (ZFC) and in a field of 100 G (FC) for LiTi$_{2-x}$V$_x$O$_4$ with $x$ = 0, 0.08, 0.12, 0.16, 0.2. The temperature hysteresis and the onset of diamagnetic response in $M/H$ diminished for $x >$ 0.12.  
Figure \ref{fig_Tc} (b) shows V concentration dependence of $T_\mathrm{c}$ and the relative volume fraction of superconductivity to $x$ = 0 at 5 K. 
The V impurities suppress the relative volume fraction faster than $T_\mathrm{c}$~\cite{SunV}.  

Low field magnetization curves were non-linear even at 300 K, but high field ones ($H >$ 0.5 T) were linear.
Thus, we defined the intrinsic magnetic susceptibility of  LiTi$_{2-x}$V$_x$O$_4$ in the normal states
by the difference in magnetization $M$ at $H$ = 4 and 5 T, $\Delta M/\Delta H$. 
The non-linear magnetization at lower fields than 0.4 T might be due to unintentional magnetic impurities (minute impurity phase). 
Figure \ref{fig_XT} shows paramagnetic susceptibility $\Delta M/\Delta H$ for $x$ = 0, 0.04 and 0.16.  
The V impurities induce Curie-Weiss like behaviors~\cite{Tsuda}.

\begin{figure}[tbp]
	\begin{center} 
		\includegraphics[width=6.8cm, clip]{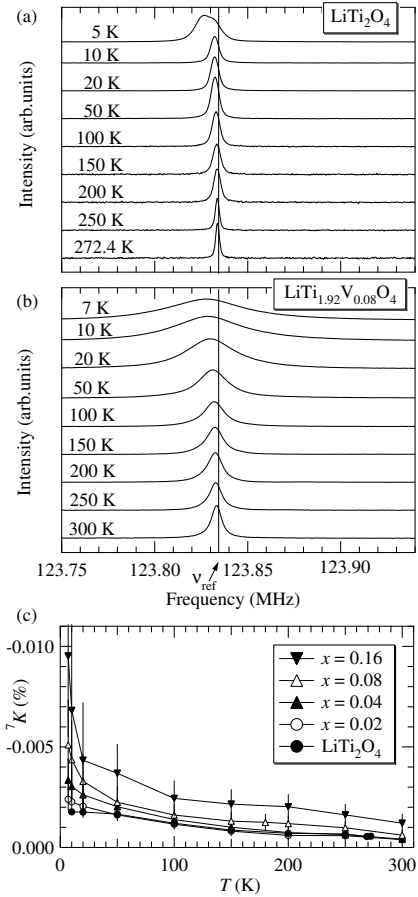}
	\end{center}
	\caption{Fourier-transformed $^{7}$Li NMR spectra of LiTi$_{2}$O$_4$ (a) and  LiTi$_{1.92}$V$_{0.08}$O$_4$ (b). The peak frequency of a $^{7}$Li NMR spectrum of LiCl$aq$ is denoted by $\nu_\mathrm{ref}$ for reference.
	(c) Temperature dependences of $^{7}$Li Knight shifts for LiTi$_{2-x}$V$_{x}$O$_4$. Note the direction of the vertical axis to the negative values. The sold lines are visual guides. 
	}
		\label{fig_FT}
\end{figure}%
Figure \ref{fig_FT} shows $^{7}$Li NMR frequency spectra as a function of temperature for $x$ = 0 (a) and $x$ = 0.08 (b). 
All the $^{7}$Li NMR spectra except $x$ = 0 below $T_\mathrm{c}$ are symmetric.
No quadrupole splits are observed. 
The peak frequencies show negative shifts and decrease as temperature is decreased. 
The linewidths are broadened by the V substitution. 
This evidences the actual substitution of the V ions for the Ti ions. 
 
Figure \ref{fig_FT}(c) shows the temperature dependences of $^{7}$Li Knight shifts  in LiTi$_{2-x}$V$_{x}$O$_4$ ($x \leq$ 0.16). The negative shifts show Curies-Weiss behaviors. The  $^{7}$Li Knight shift of Curies-Weiss sort in LiV$_{2}$O$_{4}$ was positive~\cite{Fujiwara}.

\begin{figure}[tbp]
	\begin{center}
		\includegraphics[width=6.5cm, clip]{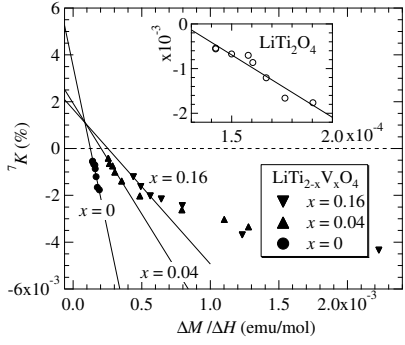}
	\end{center}
	\caption{$^{7}$Li Knight shifts $K$ are plotted against magnetic susceptibility $\chi$ defined by $\Delta M$/$\Delta H$ at $H$ = 4 and 5 T, where temperature is an implicit parameter, for LiTi$_{2-x}$V$_{x}$O$_4$ of $x$ = 0, 0.04 and 0.16. The solid lines are the results from least-squares fittings. 
	The inset shows on an expanded scale the $K-\chi$ plot of pure LiTi$_2$O$_4$.    
		}
		\label{fig_KX}
\end{figure}%
In Fig. \ref{fig_KX}, $^{7}$Li Knight shifts $K$ are plotted against magnetic susceptibility $\chi$ defined by $\Delta M$/$\Delta H$ at $H$ = 4 and 5 T, where temperature is an implicit parameter, for LiTi$_{2-x}$V$_{x}$O$_4$ with $x$ = 0, 0.04 and 0.16. 
The inset shows on an expanded scale the $K-\chi$ plot of pure LiTi$_2$O$_4$.  

The $^{7}$Li Knight shift consists of a spin shift $K_s$ and a chemical shift $\sigma$, 
\begin{equation}
^{7}K = K_s + \sigma.
\label{eq_K1}
\end{equation}
The spin shift $K_s$ is expressed by a product of a hyperfine coupling constant $A_{\rm hf}$ and a temperature dependent spin susceptibility $\chi_s(T)$ 
\begin{equation}
K_s(T) = \frac{A_{\rm hf}}{N_{\rm A}\mu_{\rm B}}\chi_s(T), 
\label{eq_K2}
\end{equation}
where $N_{\rm A}$ is the Avogadro number and $\mu_{\rm B}$ is the Bohr magneton.  

The bulk magnetic susceptibility $\chi$ is given by
\begin{equation}
\chi = \chi_s(T) + \chi_{\rm orb} + \chi_{\rm dia}, 
\label{eq_K3}
\end{equation}
where $\chi_{\rm orb}$ is the Van Vleck orbital susceptibility and  $\chi_{\rm dia}$ is the diamagnetic susceptibility of inner core electrons. 

In Fig. \ref{fig_KX}, the solid lines are the least-squares fitting results by $K(T)$ = $p\chi(T)$ + $q$ ($p$ and $q$ are the fit parameters). 
The linear relation between $^{7}K$ and $\chi$ breaks down at lower temperatures for $x$ = 0.04 and 0.16. The bulk magnetic susceptibility must include Curie components being different from 
the peak Knight shifts of the broadened $^{7}$Li NMR spectra. 

From the fitting results, the hyperfine coupling constant $^{7}A_{\rm hf}$ was estimated to be $-$ 3.12, $-$ 1.05 and $-$ 0.74 kOe/$\mu_{\rm B}$Ti for $x$ = 0, 0.04 ($T >$ 100 K) and 0.16 ($T >$ 200 K), respectively. 
In passing, the $^{7}A_{\rm hf}$ of LiV$_2$O$_4$ is positive~\cite{Fujiwara}.  
From $\chi_{\rm dia}$ = $-$ 6.28$\times$10$^{-5}$ emu/f.u.mole and tentative $\chi_{\rm orb}$ = $+$ 3.33$\times$10$^{-5}$ emu/f.u.mole~\cite{TiO2},
the chemical shift $\sigma$ was estimated to be positive 0.002 $-$ 0.005 $\%$. 
Although the magnitude of $\sigma$ depends on the choice of $\chi_{\rm orb}$,
the positive chemical shift $\sigma$ is unconventional.
\begin{figure}[htbp]
	\begin{center}
		\includegraphics[width=6.5cm, clip]{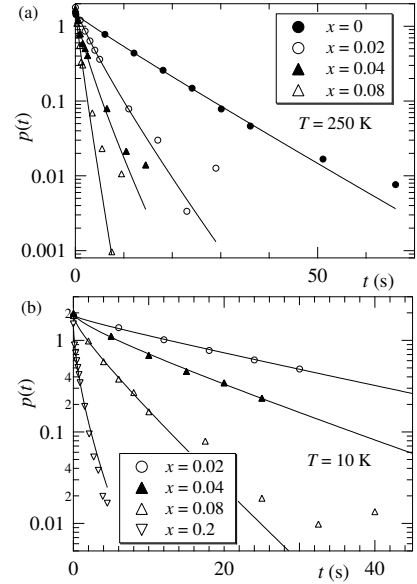}
	\end{center}
	\caption{V substitution effects on $^{7}$Li nuclear spin-lattice relaxation curves (recovery curves) in LiTi$_{2-x}$V$_x$O$_4$ ($x$ = 0$-$ 0.2). 
	}
		\label{fig_rec}
\end{figure}%
\begin{figure}[htb]
	\begin{center}
		\includegraphics[width=6.5cm, clip]{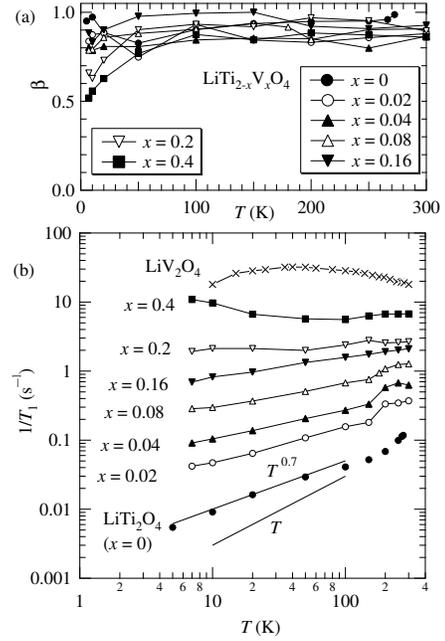}
	\end{center}
	\caption{Temperature dependences of the exponent $\beta$ of $^{7}$Li nuclear spin-lattice relaxation curves (a) and 1/$T_{1}$ (b) in LiTi$_{2-x}$V$_x$O$_4$ ($x$ = 0$-$ 0.4), and LiV$_2$O$_4$ reproduced from ref. 10. The solid curves are visual guides. 
		}
		\label{fig_T1}
\end{figure}%

Figure \ref{fig_rec} shows the $^{7}$Li nuclear spin-lattice relaxation curves (recovery curves) $p(t)\equiv 1-M(t)/M(\infty)$ ($t$ is the time after an inversion pulse to the observation pulse and $M(t)$ is the nuclear magnetization) for LiTi$_{2-x}$V$_x$O$_4$ ($x$ = 0 $-$ 0.4). 
The solid curves are the results of the least-squares fitting using a stretched exponential function
\begin{equation}
p(t) = p(0){\rm exp}[-(t/T_1)^{\beta}],
\label{eq_rec}
\end{equation}
where $p(0)$, $\beta$ and $T_1$ are the fit parameters. 
As seen in Fig. \ref{fig_rec}, the recovery curves for the V substitution are nearly single exponential functions except the low temperature for $x$ = 0.2.
The V substitution enhances 1/$T_1$ while keeping nearly the homogeneous spin-lattice relaxation. 

Figure \ref{fig_T1}(a) shows the temperature dependence of $\beta$ for $x$ = 0 $-$ 0.4. 
For $x <$ 0.2, the exponent $\beta >$ 0.8 is nearly independent of temperature, 
while for $x$ = 0.2 and 0.4, the cooling down below 50 K leads to $\beta \rightarrow$ 0.5.   
For the Li poor and rich samples, we observed $\beta \rightarrow$ 0.5 immediately when the Li deficiency and the Li substitution for Ti site~\cite{MItoh} are introduced into LiTi$_2$O$_4$. 
Thus, nearly the single exponential functions exclude the deviation of Li composition. 
The observed V impurity effect at the high magnetic field $H \sim$ 7.5 T for $x <$ 0.2 is in contrast to the conventional magnetic impurity effect on the NMR relaxation, where the non-exponential recovery curves are induced only at low fields and low temperatures ($\beta$ =1 at 300 K is reduced to 0.5 at low temperatures) and easily suppressed by the high magnetic field of $H\sim$ 7.5 T~\cite{Mc,ItohNi}.  

Figure \ref{fig_T1}(b) shows the temperature dependence of 1/$T_1$ for $x$ = 0 $-$ 0.4. 
For pure LiTi$_2$O$_4$, 1/$T_1$ shows $T^{0.7}$ dependence. 
With the V substitution, 1/$T_1$ is highly enhanced
and the temperature dependence below 100 K is changed into $T^{n}$ with $n \leq$ 0.7. 
Evidently, the enhancement of 1/$T_1$ due to the V substitution is larger than that of the magnitude of the Knight shift $K$. That is the emergence of the antiferromagnetic correlation.  

Electron correlation changes the Korringa ratio~\cite{Moriya}.
The modified Korringa relation is characterized by
\begin{align}
K(\alpha)=\frac{\gamma_e}{\gamma_{\rm n}}\frac{\hbar}{4\pi k_{\rm B}}\frac{1}{T_1TK^2_s},
\label{eq:ka}
\end{align}
where $\gamma_e$ is the electron gyromagnetic ratio and 
$\alpha$ is the exchange enhancement factor~\cite{Moriya}.
The value of $K(\alpha)$ is associated with the wave vector (${\bm q}$) dependence of a generalized spin susceptibility $\chi$($\bm q$, $\omega$), that is the ratio of the $\bm q$-averaged $\chi\prime$($\bm q$) to the uniform $\chi\prime$($\bm q$ = 0).
The ferromagnetic and antiferromagnetic $\chi\prime$($\bm q$) lead to $K(\alpha) <$ 1 and $K(\alpha) >$ 1, respectively. Incommensurately enhanced  $\chi\prime$($\bm q$) also leads to $K(\alpha) >$ 1.

One should note that the $\bm q$ dependence of the hyperfine coupling constant $A_{\rm hf}$ also plays a significant role.   
A Li site (8a) has 12 nearest neighbor Ti sites (16d).
The 12 Ti ions located at the corners of a truncated tetrahedron surround the Li ion at the center of the tetrahedron in a cubic spinel.  
Thus, the staggered magnetic fields from the Ti electrons on four sublattices may be cancelled out and masked at the Li site. 
The $\bm q$ dependent $A_{\rm hf}({\bm q})$ can work as a filter to the staggered mode.  
 
\begin{figure}[htp]
	\begin{center}
		\includegraphics[width=6.0cm, clip]{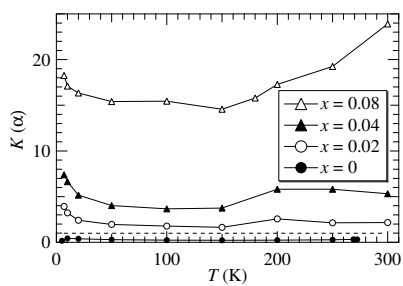}
	\end{center}
	\caption{V subsitution effect on $K(\alpha)$ of eq.(~\ref{eq:ka}). 
	The broken line indicates $K(\alpha)$ = 1.  The solid curves are visual guides.
	}
		\label{fig_Ka}
\end{figure}%
Figure~\ref{fig_Ka} shows the V substitution effect on the modified Korriga ratio $K(\alpha)$ defined by eq.(~\ref{eq:ka}). 
The V substitution changes $K(\alpha) <$ 1 for pure LiTi$_2$O$_4$ into $K(\alpha) >$ 1. 
The emergence of the antiferromagnetic correlation due to the V substitution, not a conventional quantum phase transition, was seen through the Li NMR. 
One should note that the enhanced 1/$T_1$ may be due to incommensurate magnetic correlation, 
which is not filtered by $A_{\rm hf}({\bm q})$.  

The V substitution effect is not a simple pair-breaking effect on the superconductivity
but also the change of the electron correlation effect of the 3$d$ $t_{2g}$ band 
from 3$d^{0.5}$ filling to 3$d^{1.5}$. 
The electron carrier doping via V substitution is consistent with the sign of Seebeck coefficient~\cite{Tsuda}. 

In conclusion, we observed V-induced enhancement
in $^{7}$Li nuclear spin-lattice relaxation rates in LiTi$_{2-x}$V$_x$O$_4$ ($x$ = 0 $-$ 0.4), 
which indicates the emergence of the antiferromagnetic correlation.
The V substitution for Ti ions changes the electron correlation effects by controlling the band filling from quarter electron filling 3$d^{0.5}$ to 3$d^{1.5}$. 

We thank T. Waki, H. Chudo, H. Ohta, A. Tanizawa, C. Michioka for their experimental assistance and helpful discussions. 
This work was supported in part by a Grant-in-Aid for Science Research on
Priority Area, ``Invention of Anomalous Quantum Materials," from the Ministry
of Education, Culture, Sports, Science and Technology of Japan (Grant No. 16076210)
and in part by a Grant-in-Aid for Scientific Research from the Japan Society
for the Promotion of Science (Grant No. 19350030).

\end{document}